\documentclass[preprint,prl,nofootinbib,superscriptaddress,showpacs,showkeys]{revtex4}

\usepackage{graphicx,epsfig}
\usepackage{amsfonts,amsmath,amssymb,amsthm,amscd}

\begin{document}

\title{Multiple colliding electromagnetic pulses: a way to lower
the threshold of $\mathbf{e^+e^-}$ pair production from vacuum.}

\author{S. S. Bulanov}
\affiliation{FOCUS center and Center for Ultrafast Optical Science,
University of Michigan, Ann Arbor, Michigan 48109, USA}
\affiliation{Institute of Theoretical and Experimental Physics,
Moscow 117218, Russia}

\author{V.D. Mur}
\affiliation{National Research Nuclear University MEPhI, 115409
Moscow, Russia}

\author{N.B. Narozhny}
\affiliation{National Research Nuclear University MEPhI, 115409
Moscow, Russia}

\author{J. Nees}
\affiliation{FOCUS center and Center for Ultrafast Optical Science,
University of Michigan, Ann Arbor, Michigan 48109, USA}

\author{V. S. Popov}
\affiliation{Institute of Theoretical and Experimental Physics,
Moscow 117218, Russia}

\begin{abstract}
The scheme of simultaneous multiple pulse focusing on one spot
naturally arises from the structural features of projected new laser
systems, such as ELI and HiPER. It is shown that the multiple pulse
configuration is beneficial for observing $\mathbf{e^+e^-}$ pair
production from vacuum under the action of sufficiently strong
electromagnetic fields. The field of the focused pulses is described
using a realistic three-dimensional model based on an exact solution
of the Maxwell equations. The $\mathbf{e^+e^-}$ pair production
threshold in terms of electromagnetic field energy can be
substantially lowered if, instead of one or even two colliding
pulses, multiple pulses focused on one spot are used. The multiple
pulse interaction geometry gives rise to subwavelength field
features in the focal region. These features result in the
production of extremely short $\mathbf{e^+e^-}$ bunches.
\end{abstract}

 \maketitle

One of the most profound phenomena in the Quantum Electrodynamics
(QED) of intense fields is the production of electron-positron pairs
from vacuum under the action of a strong electromagnetic (EM) field
\cite{Sauter, H&E, Schwinger,NN70}. This nonlinear phenomenon
attracts significant interest due to the fact that it lies beyond
the scope of the perturbation theory and sheds light on the
\textit{nonlinear} QED properties of the vacuum. The
$\mathbf{e^+e^-}$ production by strong EM fields in vacuum is
crucial for understanding a number of astrophysical phenomena
\cite{Ruffini_astro}. This process also places a natural physical
limit on attainable laser pulse intensity due to EM pulse energy
depletion \cite{Narozhny0406,BFP}. Moreover the process of pair
production was extensively discussed in a number of papers on the
particle formation process in high energy hadronic interaction and
the creation of quark-gluon plasmas \cite{QCD}. The
$\mathbf{e^+e^-}$ pair production process was first considered in a
static electric field, then its theoretical description was extended
to time-varying electric-type fields \cite{early papers}. Until
recently, these results were generally believed to be purely of
theoretical interest since the value of the electric field strength
needed to produce a noticeable quantity of $\mathbf{e^+e^-}$ pairs,
the the critical QED field $ E_S=m_e^2 c^3/e\hbar=1.32\times
10^{16}~\mbox{V/cm}$ (the corresponding intensity
$I_S=E_S^2/4\pi=4.65\times10^{29}$ W/cm$^2$), seemed to be
unreachable experimentally. However, the rapid development of laser
technologies promises substantial growth of peak laser intensities.
The intensity $I=2\times 10^{22}$ W/cm$^2$ is already available now
\cite{Maksimchuk} and projects to achieve $I=10^{26-28}$ W/cm$^2$
\cite{TajimaMourouBulanov,ELI,HiPER} are under way. Therefore
various aspects of $\mathbf{e^+e^-}$ pair production by focused
laser pulses are becoming urgent for experiments and are currently
gaining much attention \cite{Narozhny0406,recent papers}.

The way to obtain EM field strength close to $E_S$ in the laboratory
frame lies in generating of very short and sharply focused laser
pulses. Analytically, such pulses can be described by a realistic 3D
model developed in Ref.~\cite{NarozhnyFofanov}. Unlike the case of
spatially homogeneous time-varying electric field \cite{early
papers}, this model is based on an exact solution to the Maxwell
equations and was successfully used in \cite{Narozhny0406} for
studying the effect of $\mathbf{e^+e^-}$ pair creation by focused
circularly polarized laser pulses in vacuum. It was shown, in
particular, that the effect becomes experimentally observable at
intensities on the order of $I=10^{28}$ W/cm$^2\ll I_S$ for a single
focused pulse. This is explained by a huge value of the
pre-exponential factor in the formula for the number of created
pairs which is of the order of the ratio of the effective laser
pulse 4-volume, where pairs are effectively created, to the
characteristic Compton 4-volume. It was also shown that the
threshold intensity for the case of two head-on colliding laser
pulses is much lower and is on the order of $10^{26}$ W/cm$^2\ll
I_S$. A similar result was demonstrated recently in Ref.
\cite{Dunne}, where the superposition of a focused optical pulse
with an x-ray beam enhances the pair production.

In the present letter we use the model \cite{NarozhnyFofanov} to
consider the effect of $\mathbf{e^+e^-}$ pair creation in vacuum by
several colliding coherent linearly polarized laser pulses. Such
configurations are justified by the fact that the scheme of
simultaneous multiple pulse focusing arises naturally from the
structural features of projected new laser systems, such as ELI
\cite{ELI} and HiPER \cite{HiPER} and is implemented at NIF
\cite{NIF}. We argue that collision of four or more pulses
essentially enhances the effect of pair production as compared with
the case of a single or even two colliding pulses of the same total
input energy. The total 4-volume of the resultant field decreases
while the peak field grows. The number of created pairs depends on
the peak field exponentially while the effective laser pulse
4-volume decreases as a power. This explains the decrease of the
threshold intensity for the case of a many-pulse collision. Moreover
the interference of colliding waves generates a spotty temporal and
spatial EM field structure in the focus that leads to the generation
of ultra-short (tenths of a wavelength) electron and positron
bunches, being another way to produce ultra-short electron bunches
with intense focused EM pulses \cite{L3}.

To calculate the number of $\mathbf{e^+e^-}$ pairs produced by a
single pulse as well as by two or more colliding pulses, the fact
that the length of the formation of the pair production process is
determined by the Compton wavelength which is six orders of
magnitude shorter than the typical laser radiation wavelength,
\textit{i.e.} $\lambda \gg l_c (=3.86 \times 10^{-11} ~\mbox{cm})$
was used. At an arbitrary field point, which is characterized by the
field invariants $\mathcal{F}=(\mathbf{E}^2-\mathbf{H}^2)/2$ and
$\mathcal{G}=\mathbf{E}\mathbf{H}$, the number of pairs produced in
a unit volume per unit time can be calculated by the formula for a
constant EM field and the total number of particles produced is
calculated as the following integral over volume $V$ and time
\cite{Narozhny0406} ($\hbar=1$, $c=1$):
\begin{equation} \label{number}
N=\frac{e^2E_S^2}{4\pi^2}\int dV\int_{-\infty}^{\infty}
dt~\epsilon\eta
\coth\frac{\pi\eta}{\epsilon}\exp\left(-\frac{\pi}{\epsilon}\right).
\end{equation}
Here $ \epsilon=\mathcal{E}/E_S$, $\eta=\mathcal{H}/E_S$, and $
\left(\mathcal{E},\mathcal{H}\right)=\sqrt{\left(\mathcal{F}^2+\mathcal{G}^2\right)^{1/2}
\pm\mathcal{F}}, $ are the invariants that have the meaning of the
electric and magnetic field strengths in the reference frame where
they are parallel to each other.

In the general case, electric and magnetic fields of a focused pulse
have longitudinal components, being superpositions of two waves: the
e-wave and the h-wave that have either electric or magnetic
transverse field components respectively. There exists an exact
solution to the Maxwell equations that describes the EM field of a
linearly polarized focused pulse with focal spot radius $R$ and
Rayleigh length $L$ \cite{NarozhnyFofanov}:
\begin{equation} \label{E_e}
\begin{array}{l}
\mathbf{E}^e=iE_0 e^{-i\varphi}\left\{\left(F_1-F_2\cos
2\phi\right)~\mathbf{e}_x -F_2\sin2\phi ~\mathbf{e}_y \right\},
\\
\mathbf{H}^e=iE_0
e^{-i\varphi}\left\{\left(1-i\Delta^2\partial_\chi\right)\left[
F_2\sin2\phi ~\mathbf{e}_x \right.\right. \\ \left.\left. +
\left(F_1-F_2\cos 2\phi\right)~\mathbf{e}_y\right] +2i\Delta
\sin\phi~\partial_\xi F_1 ~\mathbf{e}_z\right\}.
\end{array}
\end{equation}
Here $x$, $y$, and $z$ are spatial coordinates, and
$\varphi=\omega(t-z)+\tilde{\varphi}$, where $\tilde{\varphi}$ is
the carrier-envelope phase, $\xi=\rho/R$, $\chi=z/L$,
$\rho=\sqrt{x^2+y^2}$, $\cos\phi=x/\rho$, $\sin\phi=y/\rho$,
$\Delta\equiv 1/\omega R=\lambda/2\pi R$, $L\equiv R/\Delta$. The
electric and magnetic fields of the h-wave are expressed via the
fields of the e-wave \cite{Narozhny2005}.

The exploited model admits different field configurations, which are
determined by two functions $F_1, F_2$. In particular, if $\Delta\ll
1$ they can be chosen in the form
\begin{equation}\label{G_b}
\begin{array}{c}
\displaystyle
F_1=(1+2i\chi)^{-2}\left(1-\frac{\xi^2}{1+2i\chi}\right )
\exp\left(-\frac{\xi^2}{1+2i\chi}\right),\\ \displaystyle
F_2=-\xi^2(1+2i\chi)^{-3}\exp\left(-\frac{\xi^2}{1+2i\chi}\right)\,,
\end{array}
\end{equation}
see Ref.~\cite{NarozhnyFofanov}. We will work with expressions
(\ref{G_b}) for functions $F_1, F_2$ throughout the paper and
consider pair production by an e-wave.

To describe a laser pulse with finite duration $\tau$ it is
necessary to introduce a temporal amplitude envelope $g((t-z)/\tau)$
and to make the following substitutions in Eqs.~(\ref{E_e})
\cite{NarozhnyFofanov} $\exp(-i\varphi) \rightarrow if'(\varphi)$,
$\Delta\exp(-i\varphi)\rightarrow \Delta f(\varphi)$, where
$f(\varphi)=g[(\varphi-\tilde{\varphi})/\omega\tau]\exp(-i\varphi)$.
It is assumed that the function
$g[(\varphi-\tilde{\varphi})/\omega\tau]=1$ at
$\varphi-\tilde{\varphi}=0$ and decreases exponentially at the
periphery of the pulse for $|\varphi|\gg\omega\tau$. In this case
the electric and magnetic fields of the model constitute an
approximate solution of the Maxwell equations having second-order
accuracy with respect to small parameters $\Delta$ and
$\Delta^\prime=1/\omega\tau$, $\Delta^\prime\lesssim\Delta \ll 1\,$.

The EM field invariants in the case of a single pulse have the
following form
\begin{equation}\label{inv1}
\begin{array}{l}
\displaystyle
\mathcal{F}_1^e=\Delta^2E_0^2\left\{\Im(F_1e^{-i\varphi})\Re[(F_{1\chi}-F_{2\chi}\cos
2\phi)e^{-i\varphi}]\right.\\
\displaystyle \left.
+\Im(F_2e^{-i\varphi})\Re[(F_{2\chi}-F_{1\chi}\cos
2\phi)e^{-i\varphi}]-2[\Re(F_{1\xi}e^{-i\varphi})]^2\sin^2\phi+O(\Delta^2)\right\}\\
\displaystyle \mathcal{G}_1^e=\Delta^2E_0^2\sin
2\phi\left\{\Im(F_2e^{-i\varphi})Re(F_{1\chi}e^{-i\varphi})
-\Im(F_1e^{-i\varphi})Re(F_{2\chi}e^{-i\varphi})+O(\Delta^2)\right\},
\end{array}
\end{equation}
where $F_{i\alpha}=\partial_\alpha F_i$, $i=1,2$ and
$\alpha=\chi,\xi$. Since $\mathcal{F}_1^e$ and $\mathcal{G}_1^e$ are
proportional to $\Delta^2E_0^2$, then the invariant fields are
proportional to $\Delta E_0$. Contrary to that, in the case of two
colliding pulses with total energy equal to the energy of a single
pulse, the invariants are no longer proportional to $\Delta$:
\begin{equation}\label{inv2}
\begin{array}{l}
\displaystyle
\mathcal{F}_2^e=2 E_0^2\left\{\left[\left|(F_1-F_2\cos2\phi)e^{i\omega z}\right|^2
+\left|F_2\sin2\phi e^{i\omega z}\right|^2\right]\sin^2\omega t\right. \\
\displaystyle \left.-\left(\Im\left[(F_1-F_2\cos2\phi)e^{i\omega
z}\right]\right)^2-\left(\Im\left[F_2\sin2\phi e^{i\omega
z}\right]\right)^2
+O(\Delta^2)\right\}\\
\displaystyle \mathcal{G}_2^e=2 E_0^2 \sin 2 \phi\left\{\Im[(F_1
F_2^{*}]\sin 2\omega t +O(\Delta^2)\right\}
\end{array}
\end{equation}
This is due to the fact that in the antinodes of standing light
waves the electric fields sum up and magnetic fields cancel each
other, \textit{i.e.} the pairs are mainly produced in the antinodes
whereas in the case of a single pulse the pairs are produced
throughout the focal 4-volume. The invariant field for two colliding
pulses is proportional to $E_0$. Since $\Delta\ll 1$, two colliding
pulses produce many more pairs than a single pulse, as was shown in
Ref. \cite{Narozhny2005} for the case of circularly polarized
pulses.

Further enhancement of the number of $\mathbf{e^+e^-}$ pairs (or
lowering the threshold intensity) can be achieved with utilization
of a configuration with multiple colliding pulses. The use of such a
set up will not only lead to focusing of a larger part of the EM
energy into a smaller volume but also to a \textit{redistribution of
the focused energy in favor of the electric field}. This will lead
to an enhancement of the invariant electric field strength and thus
to an increase in the number of pairs produced. The most beneficial
set up will be arranged if the central axes of the pulses lie in one
plane, with the pulses being linearly polarized in the direction
perpendicular to this plane. The pulses are arranged in counter
propagating pairs so that at focus their magnetic fields cancel each
other and electric fields sum up as in the antinodes of a standing
plane light wave. Then the resulting peak electric field will be
proportional to $\sqrt{n_p}$, where $n_p$ is the total number of
pulses. In the case, considered below, $n_p=8$. More pulses can be
added though with less efficiency with their central axes being at
some angle ($\theta$) to the plane where the first eight are
focused. In this case the resulting peak field in the focus will be
proportional to $(n_{p_1}+n_{p_2} \cos \theta)/\sqrt{n_p}$,
$n_{p_1}=8$ and $n_{p_1}+n_{p_2}=n_p$.

In what follows we consider a configuration where up to 24 pulses
are focused simultaneously to the same focal spot and the total EM
energy is kept constant. In Fig. 1a we show how these pulses are
focused. First eight pulses are focused in the (yz) plane in
colliding pairs along the y and z axes and along two lines ($y_+$,
$y_-$) which have angles $\pm\pi/4$ to y axis. Up to 24 pulses are
introduced by adding  pairs of pulses along the lines which do not
lie in (yz) plane and have an angle of $+\pi/4$ or $-\pi/4$ with one
of the 4 lines of pulse propagation in (yz) plane. These angles are
measured in the plane that goes through the line in (yz) plane and x
axis (Fig.1b).

\begin{figure}[ht]
\begin{tabular}{cc}
\epsfxsize3.5cm\epsffile{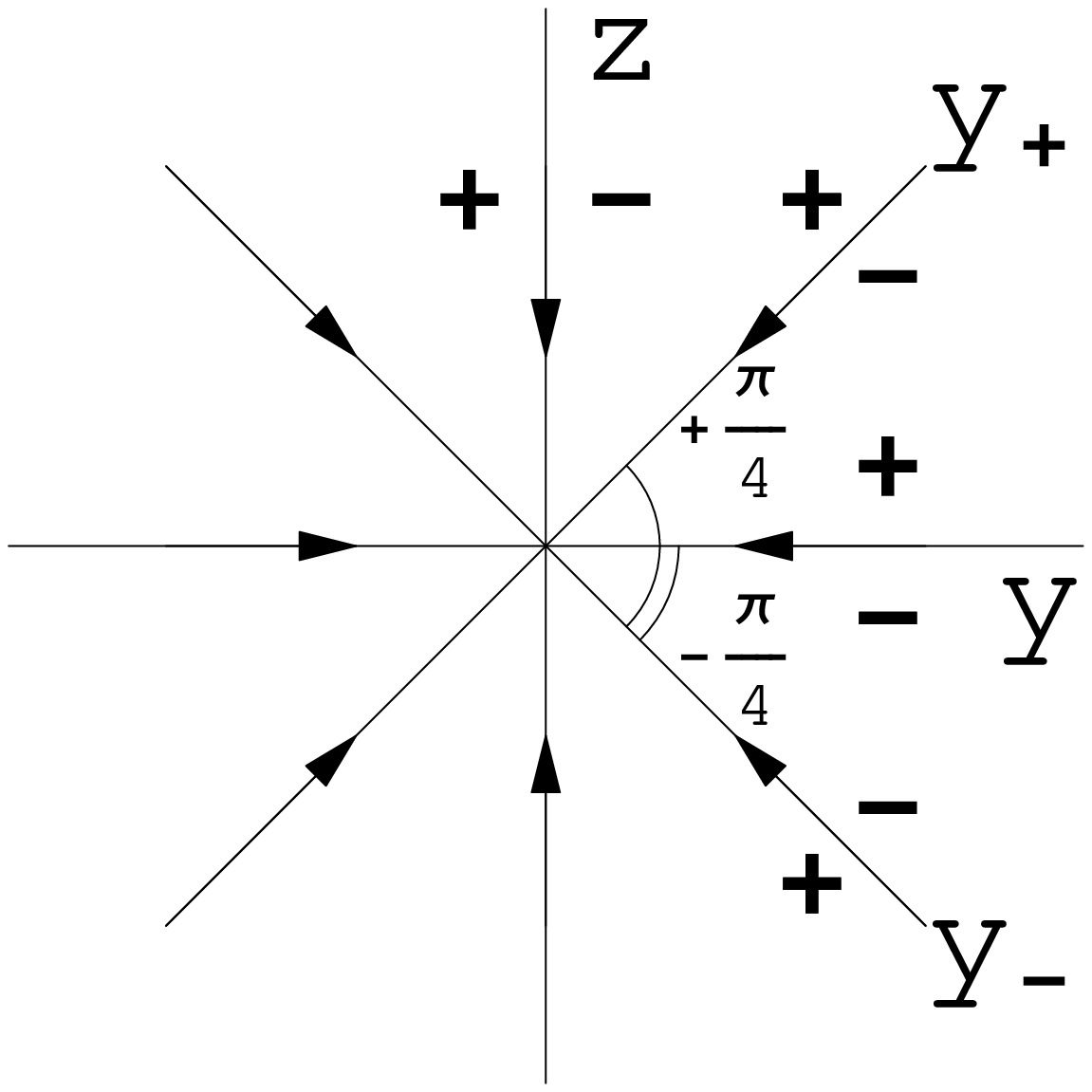} &
\epsfxsize3.5cm\epsffile{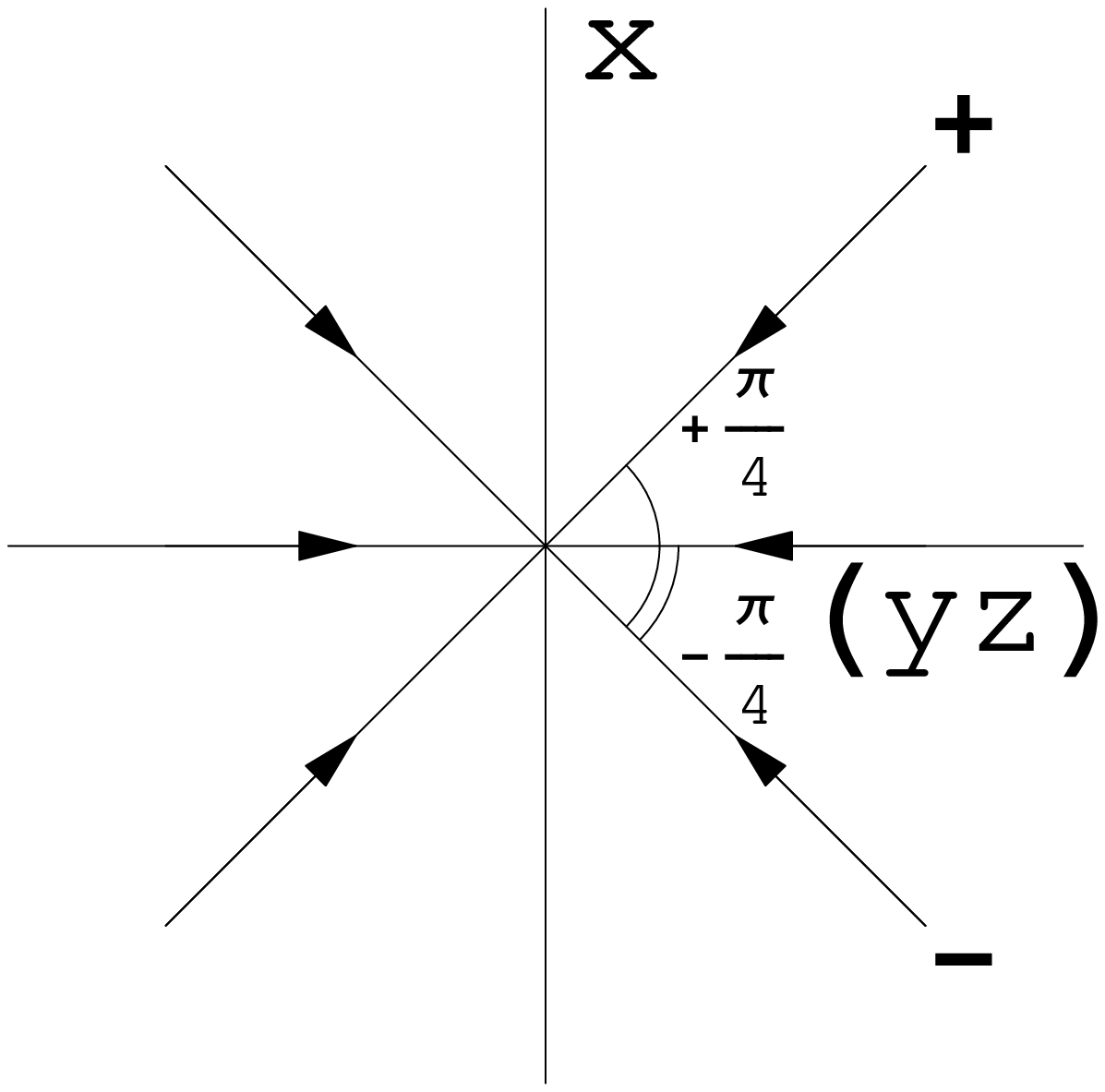}
\end{tabular} \caption{The principal scheme of multiple pulse focusing.}
\end{figure}

In Fig. 2 we present the distribution of invariant electric field in
the (xy), (yz), and (xz) planes for different numbers of pulses
focused. The duration of each pulse is 10 fs and $\Delta=0.3$. As
the number of pulses increases, so does the peak value of invariant
field. However, the volume where the invariant field is contained
shrinks, forming a spiky structure with features of about a half
wavelength in size.

\begin{figure}[ht]
\begin{tabular}{cc}
\epsfxsize11.0cm\epsffile{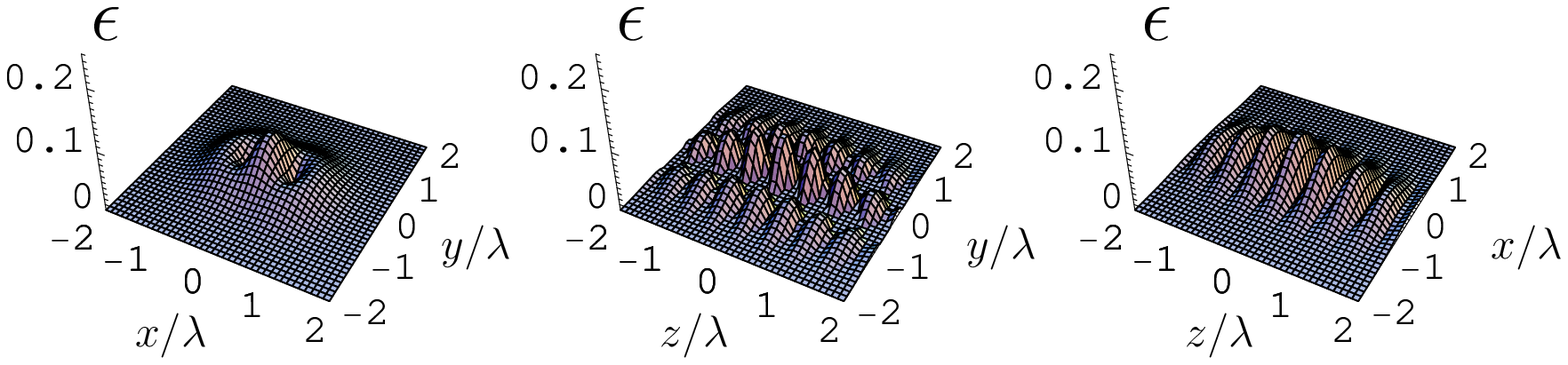}&$n_p=2$ \\
\epsfxsize11.0cm\epsffile{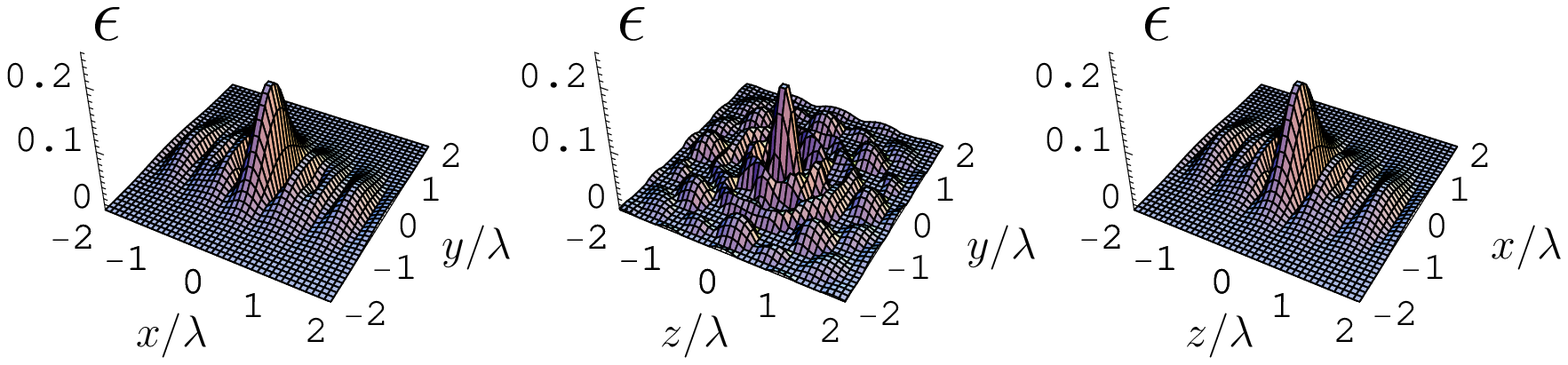}&$n_p=8$ \\
\epsfxsize11.0cm\epsffile{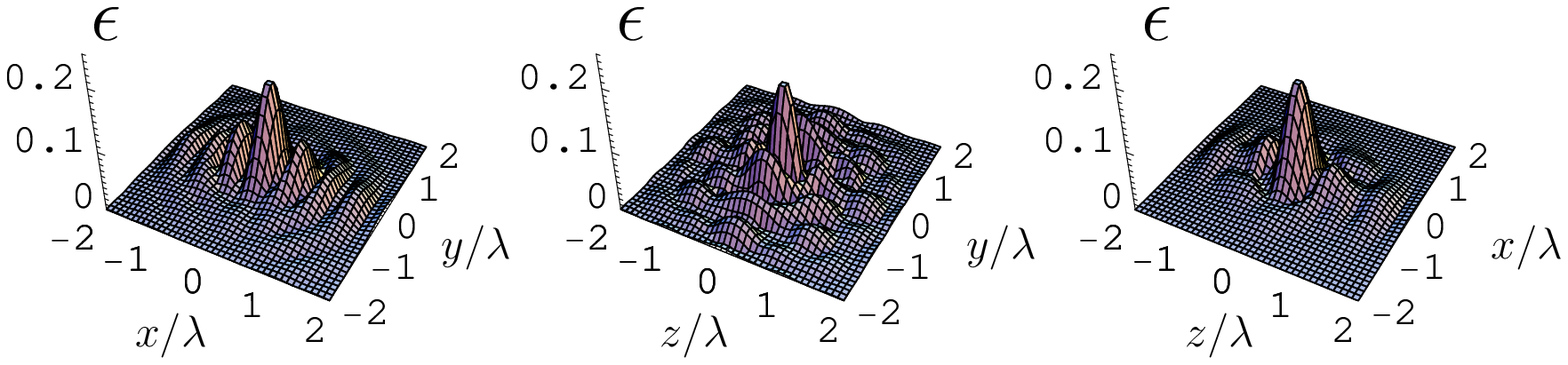}&$n_p=24$
\end{tabular} \caption{(color on-line)The distribution of invariant electric field
in (xy), (yz), and (xz) planes for 2, 8 and 24 pulses.}
\end{figure}

The time dependence of the invariant field shows a similar behavior.
We present in Figs. 3a and 3b the evolution of the invariant
electric field along x, y, and z axes for the cases of two and
twenty four pulses. It can be seen from these figures that the field
is localized in several sharp peaks. With the increase of the number
of pulses the volume where the field is localized shrinks. This will
lead to the production of very short electron and positron bunches
with characteristic duration much smaller then the radiation period.

\begin{figure}[ht]
\begin{tabular}{cc}
\epsfxsize10cm\epsffile{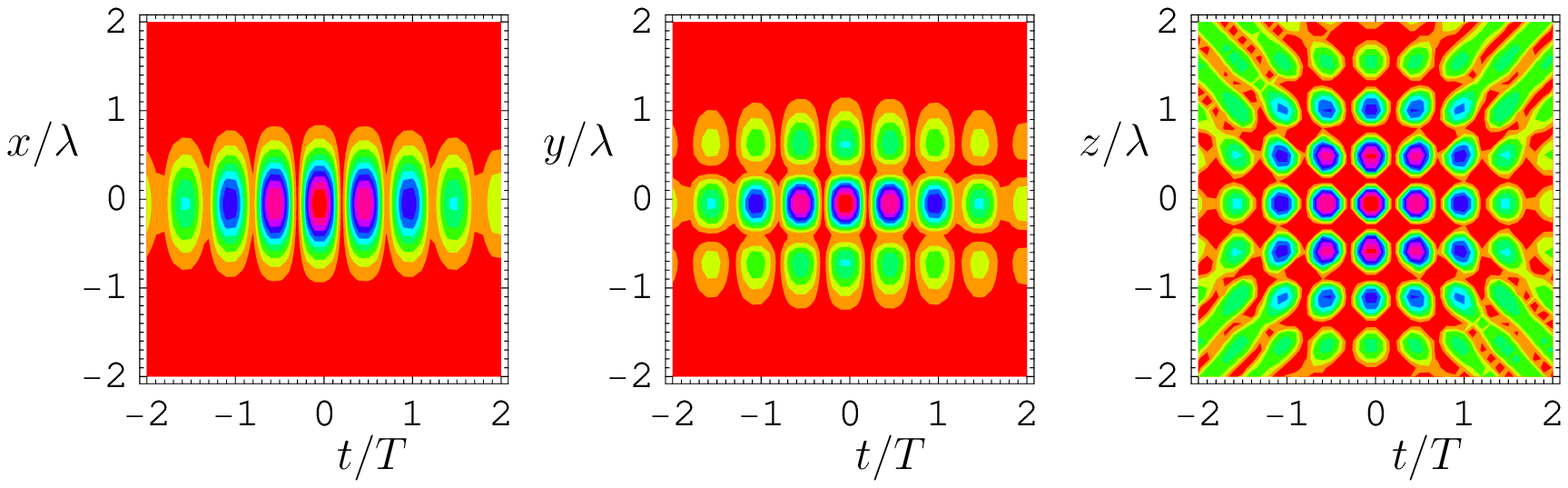} & \epsfxsize0.75cm\epsffile{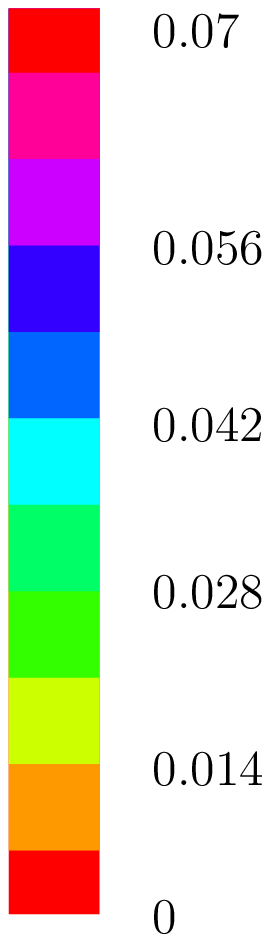} \\
\epsfxsize10cm\epsffile{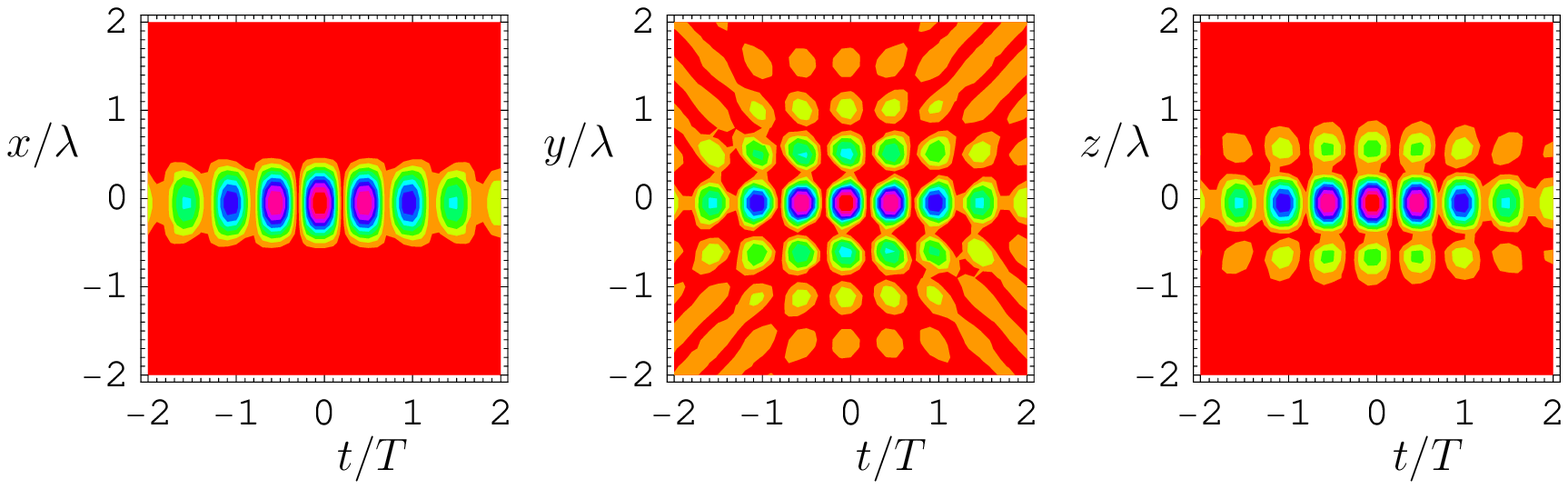} &
\epsfxsize0.65cm\epsffile{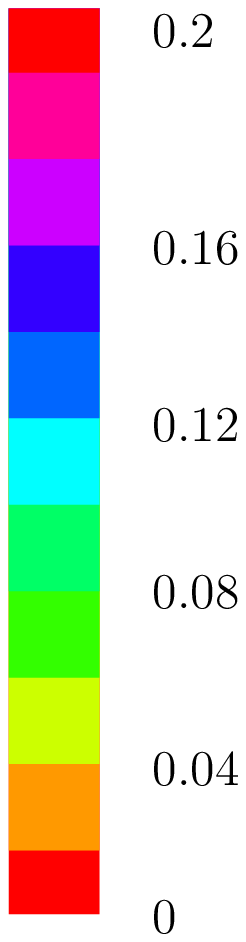}
\end{tabular} \caption{(color on-line) The evolution of invariant electric filed
along x, y, and z axes for the cases of 2 and 24 pulses.}
\end{figure}

In what follows we present the results of numerical calculations of
$e^+e^-$ pair number produced by an EM field of multiple pulses. We
use Eq. (\ref{number}) and expressions (\ref{E_e}) for the fields.
The field configuration is described in Fig. 1. Each of the pulses
has a wavelength of $\lambda=1~\mu$m, a numerical aperture
$\Delta=0.3$, a duration $\tau=10$ fs, and a focal spot of about
$\lambda$, while the total EM energy of this multipulse
configuration is a constant 10 kJ. The results are presented in
Table 1, where the number of pairs according to Eq. (\ref{number})
is shown for different numbers of pulses. Two pulses are colliding
along the z axis. Four pulses are two pairs of pulses colliding
simultaneously along z and y axes. Eight pulses are arranged in four
colliding along the y, z, $y_+$ and $y_-$ lines pairs. The 16 pulse
configuration is constructed by adding to eight in-plane pulses four
more pairs of pulses. These pulses collide along the lines that lie
in ($y_+,x$) and ($y_-,x$) planes and have an angle of $-\pi/4$ or
$+\pi/4$ with $y_+$ or $y_-$ line respectively. Twenty four pulses
represent the maximum number of pulses described in Fig. 1. We also
show the threshold energy, \textit{i.e.} the energy necessary to
produce one $e^+e^-$ pair, for the different numbers of pulses.

\begin{table}[h]
\begin{tabular}{|c|c||c|}
\hline $n$ & $N_{e^+e^-}$ at $W=10$ kJ & $W_{th}$, kJ ($N_{e^+e^-}=1$)\\
\hline
$2$ & $9.0 \times 10^{-19}$ & $40$ \\
$4$ & $3.0 \times 10^{-9}$ &  $20$ \\
$8$ & $4.0$ & $10$ \\
$16$ & $1.8 \times 10^{3}$ & $8$ \\
$24$ & $4.2 \times 10^{6}$ & $5.1$ \\
\hline
\end{tabular}
\caption{\label{<Table>}The number of $e^+e^-$ pairs ($N_{e^+e^-}$)
produced by different number of pulses (the total energy is 10 kJ
and $\Delta=0.3$); The threshold value total energy needed to
produce one $\mathbf{e^+e^-}$ pair is shown in the third column for
different numbers of pulses.}
\end{table}

According to our results pair production exceeds threshold when
eight in-plane EM pulses are simultaneously focused on one spot.
Doubling the number of pulses leads to the three orders of magnitude
increase of the number of pairs. Tripling the number of pulses makes
it possible to produce 6 orders of magnitude more pairs. The
threshold energy drops from 40 kJ for two pulses to 5.1 kJ for 24
pulses. It clearly indicates that the multiple pulse configuration
is much more favorable for $e^+e^-$ pair production than a single
pulse or even a superposition of two pulses.

As was mentioned above, the spiky temporal profile of the invariant
electric field should lead to the production of very short electron
and positron bunches with characteristic duration much smaller then
the radiation period. The duration of the central bunch can be
estimated as follows: first, we approximate the invariant electric
field as $ \epsilon=\epsilon_0\left(1-\sum\limits_i
i^2/r_i^2\right)$, $i=x,y,z,t$. Here $r_x=\lambda/2$ and
$r_y=\lambda/4$ for 2, 8 and 24 pulses, $r_z=\lambda/4,\lambda/4$
and $\lambda/2$ for 2, 8 and 24 pulses respectively and $r_t=T/4$.
Then we set $\eta=0$ and integrate (\ref{number}) over space. We get
the following expression for the dependence of the number of
produced pairs on time:
\begin{equation} \label{approx_number}
n(t)=\frac{r_x r_y r_z}{4\pi^2 l_c^4}
\left[{\epsilon}(t)\right]^{7/2}
\exp\left(-\frac{\pi}{\epsilon(t)}\right),
\end{equation}
Here $\epsilon(t)=\epsilon_0 [1-t^2/(r_t)^2]$. The duration of the
bunch at FWHM is
\begin{equation}
\Delta t=\left(\frac{\ln
2}{7/2+\pi/\epsilon_0}\right)^{1/2}\frac{T}{2}.
\end{equation}
For $\epsilon_0=0.08$ (two pulses) the electron pulse duration is
about $0.064 T$ ($190$ as for $T=3$ fs). For $\epsilon_0=0.16$
(eight pulses) the electron pulse duration is about $0.086 T$ ($260$
as for $T=3$ fs). For $\epsilon_0=0.21$ (twenty four pulses) the
electron pulse duration is about $0.097 T$ ($290$ as for $T=3$ fs).
The results of numerical calculation of bunch duration agree to this
estimate $\Delta t=0.062 T$ for 2 pulses, $\Delta t=0.089 T$ for 8
pulses, and $\Delta t=0.1 T$ for 24 pulses.

\begin{figure}[ht]
\epsfxsize5.5cm\epsffile{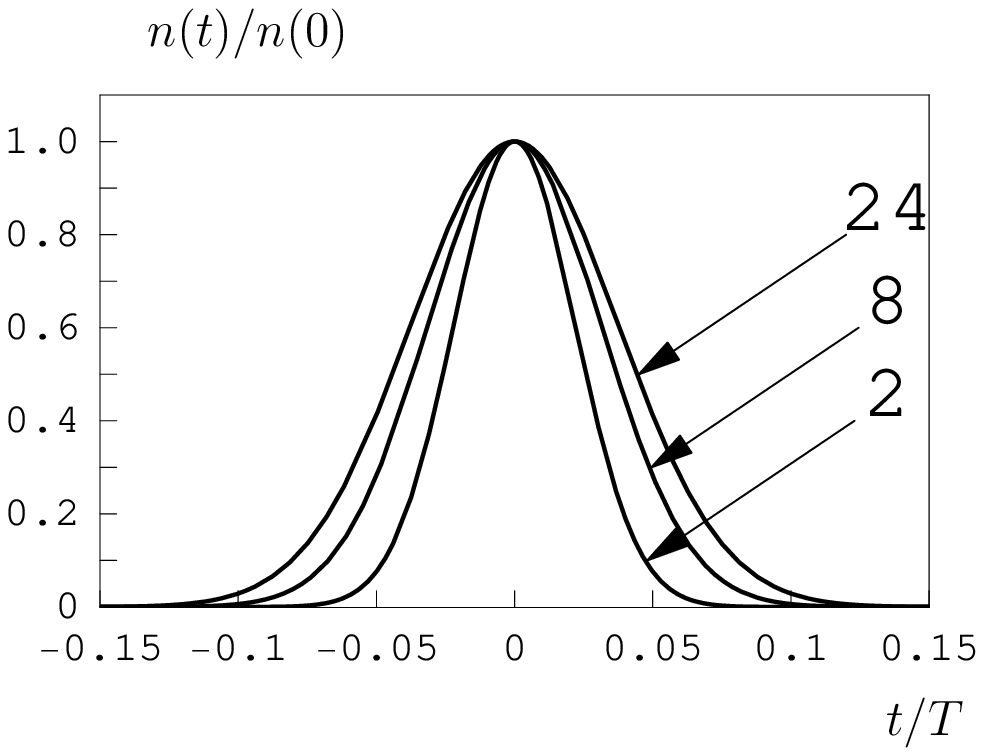} \caption{The dependence of the
number of pairs produced on time for different number of the
pulses.}
\end{figure}

In conclusion, we have showed that the simultaneous focusing of
multiple colliding pulses will lead to a significant reduction of
the threshold energy needed for the pair production to become
observable compared to the case of one or even two pulses. It is due
to the localization of the EM energy in a smaller volume and to a
redistribution of energy in favor of the electric field. According
to the results of this paper a system like ELI or HiPER with 10 kJ
of energy in 8 pulses with duration of about 10 fs will be able to
observe the $\mathbf{e^+e^-}$ pair production from vacuum by the
direct action of the EM field. And for 24 pulses, the resulting
intensity is more than adequate to produce a significant number of
$\mathbf{e^+e^-}$ pairs.

The mentioned above localization and redistribution of EM energy
leads to a structure of invariant electric field that results in the
production of ultra-short electron and positron bunches with
durations (FWHM) of about 200 as (for radiation wavelength of
1$\mu$m). This process turns out to be another way to produce
ultra-short electron/positron bunches with intense focused
electromagnetic pulses.

We would like to acknowledge fruitful discussions with G. Korn. This
work was supported by the National Science Foundation through the
Frontiers in Optical and Coherent Ultrafast Science Center at the
University of Michigan and Russian Foundation for Basic Research and
the Army Research office grant DAAD 19-03-1-0316.

\end{document}